\documentstyle[12pt]{article}
\newcommand{\bce}{\begin{center}}
\newcommand{\ece}{\end{center}}
\newcommand{\beq}{\begin{equation}}
\newcommand{\eeq}{\end{equation}}
\newcommand{\bea}{\vspace{0.25cm}\begin{eqnarray}}
\newcommand{\eea}{\end{eqnarray}}

\newcommand{\ba}{\begin{array}}
\newcommand{\ea}{\end{array}}

\newcommand{\doublespace}{
    \renewcommand{\baselinestretch}{1.6}\large\normalsize}

\def\lsim{\mathrel{\rlap{\lower4pt\hbox{\hskip1pt$\sim$}}
    \raise1pt\hbox{$<$}}}	  
\def\gsim{\mathrel{\rlap{\lower4pt\hbox{\hskip1pt$\sim$}}
    \raise1pt\hbox{$>$}}}	  

\def\Pom{{\bf I\!P}}

\def\lsim{\mathrel{\rlap{\lower4pt\hbox{\hskip1pt$\sim$}}
    \raise1pt\hbox{$<$}}}         
\def\gsim{\mathrel{\rlap{\lower4pt\hbox{\hskip1pt$\sim$}}
    \raise1pt\hbox{$>$}}}         

\def\Pom{{\bf I\!P}}

  \advance\oddsidemargin  by -1in
  \advance\evensidemargin by -1in
  \advance\topmargin      by -0.5in
\def\lsim{\mathrel{\rlap{\lower4pt\hbox{\hskip1pt$\sim$}}
    \raise1pt\hbox{$<$}}}         
\def\gsim{\mathrel{\rlap{\lower4pt\hbox{\hskip1pt$\sim$}}
    \raise1pt\hbox{$>$}}}         

\def\Pom{{\bf I\!P}}
\def\beq{\begin{equation}}
\def\endeq{\end{equation}}
\def\arr{\begin{eqnarray}}
\def\endarr{\end{eqnarray}}
\makeindex

\doublespace

\begin{document}


\phantom{.}{\bf \Large \hspace{9.6cm} KFA-IKP(Th)-1994-17\\
\phantom{.}\hspace{11.3cm}20 May 1994\vspace{0.4cm}\\ }

\begin{center}
{\bf\sl \huge Scanning the BFKL pomeron in elastic production
of vector mesons at HERA}
\vspace{0.4cm}\\
{\bf \large
J.Nemchik$^{a,b}$, N.N.~Nikolaev$^{b,c}$, and B.G.~Zakharov$^{c}$
\bigskip\\}
{\it
$^{a}$Institute of Experimental Physics, Slovak Academy of Sciences,\\
Watsonova 47, 04353 Kosice, Slovak Republik
\medskip\\
$^{b}$IKP(Theorie), KFA J{\"u}lich, 5170 J{\"u}lich, Germany
\medskip\\
$^{c}$L. D. Landau Institute for Theoretical Physics, GSP-1,
117940, \\
ul. Kosygina 2, Moscow 117334, Russia.
\vspace{1.0cm}\\ }
{\Large
Abstract}\\
\end{center}
Elastic production of vector mesons $\gamma^{*}\,N\rightarrow
V\,N$  is the pomeron-exchange dominated diffractive reaction
with much potential of probing the BFKL pomeron. The BFKL pomeron
can conveniently be described in terms of the dipole
cross section which
is a solution of the generalized BFKL equation. In this paper
we discuss, how the energy and $Q^{2}$ dependence of elastic
production
of vector mesons at HERA will allow scanning the dipole cross
section as a function of dipole size $r$. We show that
determinaton of the intercept of the BFKL pomeron requires
measuring the $\rho^{0}$ and $J/\Psi$ production at $Q^{2} \sim
(100-200)$GeV$^{2}$ and/or the quasireal photoproduction of
the $\Upsilon$.
We present predictions for the effective
intercept in the kinematic range of the forthcoming HERA
experiments, which can shed much light on the nonperturbative
component of the pomeron.
 \bigskip\\

\begin{center}
E-mail: kph154@zam001.zam.kfa-juelich.de
\end{center}

\pagebreak

Elastic real and virtual photoproduction of vector mesons
\beq
\gamma^{*}p\rightarrow Vp\,,~~~~~V=\rho^{0},\,\omega^{0},\,
\phi^{0},\,J/\Psi,\,\Upsilon\,...,
\label{eq:1}
\endeq
is the typical diffractive reaction, dominated by pomeron
exchange. Determination of parameters of the BFKL pomeron [1],
in particular of its intercept $\Delta_{\Pom}$, is an outstanding
problem for the HERA experiments on
small $x$, and exploring the potential of reaction
(1) is of great importance. In [2-4] we developed a novel
approach to the BFKL pomeron in terms of the dipole cross
section $\sigma(\xi,r)$ which satisfies the generalized
BFKL equation.
Here $r$ is the size of the color dipole,
$\xi = \log[2m_{p}\nu/(Q^{2}+m_{V}^{2})]$ is the rapidity,
$m_{V}$ is the vector meson mass, $Q^{2}$ is the virtuality
of the photon and $\nu$ its energy in the proton rest frame.
At very high $\nu$ and/or very small Bjorken
variable $x$ one expects
\beq
\sigma(\xi,r)=\sigma_{\Pom}(r)\exp(\Delta_{\Pom}\xi)
\label{eq:2}
\endeq
and
\beq
F_{2}(x,Q^{2}) \propto \exp(\Delta_{\Pom}\xi)\sim
\left({1\over x}\right)^{\Delta_{\Pom}}
\label{eq:3}
\endeq
with the intercept $\Delta_{\Pom}$ which does not
depend on $Q^{2}$, the real photoproduction $Q^{2}=0$ not
excepted. Understanding the onset of the BFKL regime
(\ref{eq:2},\ref{eq:3}) is the most pressing issue, because
experimentally a steep rise of $F_{2}(x,Q^{2})$ at HERA
[5] is accompanied by a much slower rise
of the real photoabsorption cross section [6].

What makes reactions (1) exceptionally important is the scanning
phenomenon discovered in [7-11], by which elastic production
amplitude probes the dipole cross section at the scanning radius
\beq
r_{S} \approx {C \over \sqrt{m_{V}^{2}+Q^{2}}}\, .
\label{eq:4}
\endeq
The scale parameter $C$ is rather large , and in [11] we gave an
estimate $C\approx 6$. Changing $Q^{2}$ and the mass of the produced
vector meson, one can probe $\sigma(\xi,r)$ and measure the effective
intercept $\Delta_{eff}(\xi,r)
={\partial\log\sigma(\xi,r)/ \partial\xi}
$ in a very broad range of dipole
size $r$, from the nonperturbative regime of $r \gsim 1$f~ down
to the perturbative regime of $r \ll 1$f. The HERA experiments
are already amassing the data on production of vector mesons, and
the purpose of this communication is to present the detailed
predictions for reactions (\ref{eq:1}) from the approach
to the BFKL pomeron developed in [2-4,12-15].

In order to set up the framework, we briefly review
properties of the dipole cross section. In [3,4,12,13]  we
studied solutions of the generalized BFKL equation in the realistic
model with the finite correlation radius $R_{c}$ for perturbative
gluons and the running QCD coupling $\alpha_{S}(r)$, which freezes
at large distances $\alpha_{S}(r \geq R_{f})=\alpha_{S}^{(fr)}=0.8$.
With the $R_{c}=0.3f$ as suggested by the lattice QCD studies,
$\Delta_{\Pom}=0.4$ [4,12]. The major findings are:

Firstly, the onset of the asymptotic behaviour
(\ref{eq:2},\ref{eq:3}) was found to be extremely slow. Namely,
the effective intercept
$
\Delta_{eff}(\xi,r)
$
varies quite strongly in the range of $r$ and $\xi$ relevant to
the HERA experiments on DIS. The usually discussed structure
function $F_{2}(x,Q^{2})$ receives contributions from a broad
range of $r$ and, for this reason, its $x$ dependence does not
allow a realiable experimental determination of the intercept
$\Delta_{\Pom}$.

Secondly, by the diffusion property of the BFKL kernel
[1,4,12], at
asymptotic energies the behaviour (\ref{eq:2}) takes over at all
values of $r$, including $r > R_{c}$, thus
unifying the asymptotic energy
dependence of the bare pomeron-exchange contribution to hadronic
scattering, the real photoproduction and deep inelastic scattering.
Furthermore, by the same diffusion property, the intercept
$\Delta_{\Pom}$ is almost entirely controlled by the region
of $r\sim R_{c}$ [4,12].

Thirdly, there happens to exist a "magic" size $r_{\Delta}\approx
{1\over2} R_{c}$, at which the
precocious asymptotic behaviour $\Delta_{eff}(\xi,r) \approx
\Delta_{\Pom}$ will persist over the whole range of $x$ at HERA.
In [4,15] we showed how one can zoom at $r\sim r_{\Delta}$,
measuring at HERA the charm structure function at $Q^{2}\lsim 10$
GeV$^{2}$, the longitudinal structure function $F_{L}(x,Q^{2})$
at $Q^{2} \sim (10-40)$GeV$^{2}$ and the scaling violations
$\partial F_{T}(x,Q^{2})/ \partial \log Q^{2}$ at
$Q^{2} \sim (2-10)$GeV$^{2}$.

Fourthly, the BFKL pomeron describes the exchange by perturbative
gluons with the typical interaction radius $\sim R_{c}$.
At low
energy and large dipole size $r\gg R_{c}$, the nonperturbative
scattering mechanism can take over. The rise of
the BFKL perturbative cross
section is driven by the rising multiplicity of perturbative gluons
in the high-energy photon (hadron), for the nonperturbative
mechanism one can expect the (approximately)
energy-independent contribution
$\sigma^{(npt)}(r)$ to the dipole cross section. A large contribution
from the $\sigma^{(npt)}(r)$ reconciles the steep rise of
$F_{2}(x,Q^{2})$ [5] with a much weaker energy dependence of the
real photoabsorption cross section [6]. As we shall discuss below,
the scanning phenomenon allows to study in much detail a
transition between the nonperturbative and perturbative regimes,
which makes reactions (1) particularly important.

The origin of the scanning phenomenon is as follows:
In the dipole-cross section representation, the amplitude of
the forward production $\gamma^{*}\,p \rightarrow V\,N$ of the
(T) transverse and (L) longitudinally polarized vector mesons
reads [7-11,16,17]
\arr
M_{T}(\xi,Q^{2})={N_{c}C_{V}\sqrt{4\pi\alpha_{em}} \over (2\pi)^{2}}
\cdot~~~~~~~~~~~~~~~~~~~~~~~~~~~~~~~~~
\nonumber \\
\cdot \int d^{2}\vec{r} \sigma(\xi,r)
\int_{0}^{1}{dz \over z(1-z)}\left\{
m_{q}^{2}
K_{0}(\varepsilon r)
\phi(r,z)-
[z^{2}+(1-z)^{2}]\varepsilon K_{1}(\varepsilon r)\partial_{r}
\phi(z,r)\right\}\nonumber \\
 =
{1 \over (m_{V}^{2}+Q^{2})^{2}}
\int {dr^{2} \over r^{2}} {\sigma(\xi,r) \over r^{2}}
W_{T}(Q^{2},r^{2})
\label{eq:5}
\endarr
\arr
M_{L}(\xi,Q^{2})={N_{c}C_{V}\sqrt{4\pi\alpha_{em}} \over (2\pi)^{2}}
{2\sqrt{Q^{2}} \over m_{V}}
\cdot~~~~~~~~~~~~~~~~~~~~~~~~~~~~~~~~~
 \nonumber \\
\cdot \int d^{2}\vec{r} \sigma(\xi,r)
\int_{0}^{1}dz \left\{
[m_{q}^{2}+z(1-z)m_{V}^{2}]
K_{0}(\varepsilon r)
\phi(r,z)-
\varepsilon K_{1}(\varepsilon r)\partial_{r}
\phi(z,r)\right\}\nonumber \\
 =
{1 \over (m_{V}^{2}+Q^{2})^{2}}
{2\sqrt{Q^{2}} \over m_{V}}
\int {dr^{2} \over r^{2}} {\sigma(\xi,r) \over r^{2}}
W_{L}(Q^{2},r^{2})
\label{eq:6}
\endarr
Here $N_{c}=3$ is the number of colours,
$C_{V}={1\over \sqrt{2}},\,{1\over 3\sqrt{2}},\,{1\over 3},\,
{2\over 3}~~$ for the
$\rho^{0},\,\omega_{0},\,\phi^{0},\, J/\Psi$ production,
respectively,
 $K_{0,1}(x)$ is the modified Bessel function,
\beq
\varepsilon^{2}=m_{q}^{2} + z(1-z)Q^{2}\, ,
\label{eq:7}
\endeq
$\phi(z,r)$ is the lightcone wave function of the $q\bar{q}$
Fock state of the vector meson in the mixed
$(z,\vec{r})$-representation, where $\vec{r}$ is the transverse
separation of the quark and antiquark and $z$ is a fraction of
the lightcone momentum carried by the quark.
 The normalization
is such that
\beq
{ N_{c} \over 2\pi}\int d^{2}\vec{r} \int_{0}^{1}
{dz \over z(1-z)}\left\{m_{q}^{2} \phi^{2}(r,z)+
[z^{2}+(1-z)^{2}][\partial_{r}\phi(r,z)]^{2}\right\}
=1 \, .
\label{eq:8}
\endeq
The dipole cross section can be related to the gluon structure
function of the target proton $G(x,q^{2})$ [2,17,18]
\beq
\sigma(\xi,r) = {\pi^{2} \over 3}r^{2}\alpha_{S}(r)
G(x=\exp(\xi), q^{2}={A\over r^{2}})
\propto r^{2}
\left[{1\over \alpha_{S}(r)}\right]^{\gamma-1}
\exp(\Delta_{\Pom}\xi)\, ,
\label{eq:9}
\endeq
where $\gamma = 4/3\Delta_{\Pom}$  [14] and
$A\approx 10$ (the emergence of this large numerical factor
is explained in [13]).
The most important property of the  dipole cross section
is the color transparency driven dependence
$\propto r^{2}$ at small $r$.
Because $K_{0,1}(x) \propto \exp(-x)$ at large $x$, and
the wave function of the vector meson is smooth, the amplitudes
(\ref{eq:5},\ref{eq:6}) will be dominated by the contribution
from $r\approx 3/\varepsilon$, which in the nonrelativistic
approximation of $m_{V}\sim 2m_{q}$ and $z\sim {1\over 2}$
leads to the scanning radius (\ref{eq:4})  and the estimate
$C\approx 6$. When the scanning radius is small, $r_{S}\lsim R_{V}$,
where $R_{V}$ is the vector meson radius, the amplitudes
(\ref{eq:6},\ref{eq:7}) can be evaluated as
\beq
M_{T} \propto r_{S}^{2}\sigma(\xi,r_{S}) \propto {1 \over
(m_{V}^{2}+Q^{2})^{2}} \, ,
\label{eq:10}
\endeq
\beq
M_{L} \propto
{\sqrt{Q^{2}}\over m_{V}}
 r_{S}^{2}\sigma(\xi,r_{S}) \propto
{\sqrt{Q^{2}}\over m_{V}}
 {1 \over
(m_{V}^{2}+Q^{2})^{2}} \propto
{\sqrt{Q^{2}}\over m_{V}} M_{T}\, ,
\label{eq:11}
\endeq
which must be contrasted to prediction of the
vector dominance model $M_{T} \propto 1/(m_{V}^{2}+Q^{2})$.
This departure from the vector dominance comes entirely from
colour transparency property of the dipole cross section [11].
Production of the longitudinally polarized vector mesons
dominates at $Q^{2}\gg m_{V}^{2}$.

The scanning property is quantified by weight function
$W_{T,L}(Q^{2},r^{2})$ in Eqs.~(\ref{eq:5},\ref{eq:6}), which
we introduced following [13].
The $W_{T,L}(Q^{2},r^{2})$
are sharply
peaked (Fig.~1) and, because $\sigma(\xi,r)/r^{2}$ is a slow
function of $r$, see Eq.~(\ref{eq:9}), the position of the
peak gives the scanning radius $r_{S}$. At sufficiently large
$Q^{2}$ such that $r_{S} \lsim R_{V}$, which typically means
$Q^{2} \gg m_{V}^{2}$, the position of the peak
in the variable $y=r^{2}(Q^{2}+m_{V}^{2})$ very weakly depends
on $Q^{2}$. For production of the longitudinally polarized vector
mesons, the peak corresponds to $C_{L}(J/\Psi)\sim 7$
and $C_{L}(\rho^{0})\sim 8$. For the transversely polarized
$\rho^{0}$ mesons, with rising $Q^{2}$
the peak is slowly drifting towards
$C_{T}\sim 12$ at $Q^{2} \sim 100$GeV$^{2}$,
which can be understood as follows [11]:
At a fixed $z$, the scanning radius $r_{S}\approx 3/\varepsilon$.
For the longitudinal photons, the wave function is peaked at
$z\sim {1\over 2}$, hence the estimate $C\approx 6$ is rather
accurate (for the related discussion of the longitudinal
structure function see [2,17,20]). In the opposite to that,
the transverse structure function receives
significant contribution from the very asymmetric pairs
with $z,(1-z) \sim m_{q}^{2}/Q^{2}$, which have a large, hadronic
transverse size $r\sim 1/m_{q}$ [2,17]. Because of the contribution
of such asymmetric pairs, the effective scanning radius for the
transversely polarized vector mesons decreases with $Q^{2}$
less rapidly than given by the simple estimate (\ref{eq:4}).
The results shown in Fig.~1, were obtained with the relativized
wave function of vector mesons, which has the correct short
distance behaviour driven by hard gluon exchange [16]. Notice,
that because of very large numerical value of $C_{T,L}$, the scanning
radius $r_{S}$ Eq.~(\ref{eq:4}) remains large, and the relativistic
effects in wave functions are still marginal, in a very broad range
of $Q^{2}$. For instance, for the $J/\Psi$
we find $C_{T}\approx C_{L}$ and very close to the nonrelativistic
estimate $C_{T,L}\approx 6$ even at $Q^{2}\gg m_{V}^{2}$.

At small $Q^{2}$, the scanning radius $r_{S}$ is large, and
one probes the dipole cross section in the nonperturbative
domain of $r$. In Fig.~2 we present the decomposition of the
dipole cross section into the nonperturbative component
$\sigma^{(npt)}(r)$ and the $\sigma^{(pt)}(\xi,r)$ which is a
solution of the perturbative BFKL equation\footnote{This Fig.~2
updates Fig.~1 of Ref.~[14], in which the mistake was made in
plotting $\sigma^{(pt)}(\xi,r)$. This plotting mistake does not
affect any of predictions made in [14], though. Our choice of
$\sigma^{(npt)}(r)$ in this paper is slightly different from that
in [14].}
:
\beq
\sigma(\xi,r)=
\sigma^{(npt)}(r)+\sigma^{(pt)}(\xi,r)
\label{eq:12}
\endeq
This additivity of the bare pomeron cross sections and/or of the
eikonal functions when the unitarization is considered,
is the simplest
assumption, and more refined treatment will be necessary in an
analysis of the future high-precision data. The shape of the
nonperturbative cross section is the largest unknown in
the problem, and is mostly driven by our analysis of the
$Q^{2}$ dependence of the photoproduction of vector mesons [8-11]
and of structure functions at moderate $Q^{2}$ [18],
which are well described by the dipole cross section calculated
in [17]. Therefore, the $\sigma^{(npt)}(r)$ is tuned so as
to reproduce the dipole cross section of ref.~[17] at
$\xi=\xi_{0}= -\log 0.03$, which we take as the starting point
for the BFKL evolution.
In Fig.~2 we show how the $\sigma^{(pt)}(\xi,r)$ evolves with
the rapidity $\xi$. At small $r$, the total dipole cross
section is dominated by the BFKL cross section
$\sigma^{(pt)}(\xi,r)$. In this region, because of the
so-called double logarithmic effects, the effective intercept
$\Delta_{eff}(\xi,r) > \Delta_{\Pom}$ [3,4]. At large $r$,
$\Delta_{eff}(\xi,r) \ll \Delta_{\Pom}$ because of the large
contribution from the nonperturbative cross section
$\sigma^{(npt)}(r)$. The analysis [4,13] has shown that there exists
a magic radius $r\sim r_{\Delta}$, at which
$\Delta_{eff}(\xi,r) \approx \Delta_{\Pom}$ starting already
at moderate $\xi$. At this value of $r$, the nonperturbative
cross section is small, and zooming at $r\sim r_{\Delta}$ shall
allow direct measurement of $\Delta_{\Pom}$ at HERA.

The above discussion refers to the bare pomeron cross section.
With the conventional Gaussian parametrization of the elastic
scattering peak, $d\sigma_{el}/dt \propto \exp(-B|t|)$, where
$|t|$ is the momentum transfer squared, the profile function
of elastic scattering takes the form
\beq
\Gamma(b)={\sigma_{tot} \over 4\pi B}
\exp\left(-{b^{2} \over 2B}\right)  \, ,
\label{eq:13}
\endeq
where $b$ is the impact parameter. The rise of the bare pomeron
cross section will eventually conflict the
$s$-channel unitarity bound
$\Gamma(b) \leq 1$. Unique solution of the unitarization problem
is lacking; for the crude estimate we apply the ${\cal K}$-matrix
unitraization
\beq
\Gamma(b)={\Gamma_{0}(b)\over 1+\Gamma_{0}(b)}\,.
\label{eq:14}
\endeq
When the rising strength of the bare pomeron interaction
$\Gamma_{0}(b) \gg 1$, the unitraized profile function tends to
the black disc limit $\Gamma(b)\rightarrow 1$.
The ${\cal K}$-matrix unitarization leads to a particularly simple
form of the unitarized total cross section [19]
\beq
\sigma^{(U)}(\xi,r)=
4\pi B\log\left(1+{\sigma(\xi,r) \over 4\pi B}\right)\, .
\label{eq:15}
\endeq
For a crude evaluation of the unitarization effects, we take the
energy-independent diffraction slope $B=10$GeV$^{-2}$.
Evidently, the unitarization effects only become important when
$\sigma(\xi,r) \gsim 4\pi B  \sim 50 mb$, i.e., either at large
$r$ and/or at very high energy. As we shall see below, in the
kinematical range of the HERA experiments, unitarization effects
are marginal.

Our predictions for the energy dependence of different
photoproduction observables are shown in Fig.~3,4. In Fig.~3
we show the real photoabsorption cross section. It was
calculated using the wave function of the photon derived in
[17], assuming the effective quark mass $m_{u,d}=0.15$GeV, which
is the sole parameter in the quark wave function. Only the
nonperturbative contribution to $\sigma_{tot}(\gamma p)$ is
sensitive to this parameter.
For the comparison purposes, we also show our predictions
for $\sigma_{tot}(\rho^{0}p)$, which is close to
$\sigma_{tot}(\pi N)$ and has a rise
consistent with the observed trend of the hadronic
total cross sections (for a recent review and
high-energy extrapolations see [22]). We attribute the slow rise
of $\sigma_{tot}(\gamma p)$ and $\sigma_{tot}(\rho^{0}p)$
to the large contribution from the nonperturbative large-size
cross section (for the early  discussion of such a scenario see
[23]). The effect of the unitarization slowly rises with energy.
It predominantly affects the nonperturbative cross section and,
given the uncertanty in the absolute normalization of
the $\sigma^{(npt)}(r)$, at small energies it can be compensated
for by upwards renormalization of the input $\sigma^{(npt)}(r)$.
The energy dependent part of the unitarization correction is,
however, the genuine effect and brings our results for
$\sigma_{tot}(\gamma p)$ to a better agreement with the
experiment [6]. The simple ${\cal K}$-matrix unitarization, as
well as the eikonal unitarization, lacks the so-called
triple-pomeron contribution, which enhances the effect of
unitarization at high energies ([2,19] and references therein) and
can tame the somewhat too rapid a growth of
$\sigma_{tot}(\gamma p)$, a description
of the low energy data [21] also can be augmented,
by adding to the nonperturbative
cross-section the Regge-behaving $\propto 1/\sqrt{\nu}$
terms [22]. However, the
purpose of the present analysis is understanding the gross features
of the BFKL phenomenology rather than fitting the low-$Q^{2}$
experimental
data.
Recently, there was much discussion on the
perturbative QCD contribution to the real photoabsorption cross
section $\sigma_{tot}(\gamma p)$, in which the eikonal is
evaluated with the BFKL cross section substituted for the
inclusive minijet cross section ([24] and references therein)
which is incorrect (for the related criticism see [25]).

In Fig.~4 we show how the unitarization affects the energy
dependence of $d\sigma /dt |_{t=0}$ for the real photoproduction
of the $\rho^{0}$ and $J/\Psi$. In the former case the scanning
radius is large, $r_{S} \gsim 1$f, in the latter case
$r_{S}\sim 0.4$f. We also show the unitarization effect for the
virtual photoproduction of the longitudinally polarized $\rho^{0}$
meson at $Q^{2} = 120$GeV$^{2}$, appropriate for the scanning
radius $r_{S}\sim r_{\Delta}\sim 0.15$f. Notice a rapid decrease
of the unitarization correction with the decrease of the scanning
radius $r_{S}$.

For a crude estimate of elastic production cross section we can
take $B(\rho^{0}) \approx 10$GeV$^{2}$ [6]
and $B(J/\Psi)\approx 4$GeV$^{2}$
[26]. For the real photoproduction of the $\rho^{0}$ at
$\sqrt{s}=\sqrt{2m_{p}\nu}=200$GeV, we find
$\sigma(\gamma p\rightarrow \rho^{0} p)/\sigma_{tot}(\gamma p)=
0.085$ (for the bare, non-unitarized, cross sections this ratio
equals 0.135 ), in good agreement with the ZEUS determination
$0.1\pm 0.04$ [6]. For the real photoproduction of the $J/\Psi$
at $\nu=150$GeV we find $\sigma(\gamma p \rightarrow J/\Psi p)
\sim 16$nb, which agrees with the E687 result of $17.9\pm 4.0
$nb at $\nu=177$GeV and $9.8\pm 2.9$nb at $\nu=121$GeV [26]. The
rise of the $J/\Psi$ production cross section with energy observed
by E687 is also consistent with our prediction in Fig.~4. Our
estimate for the real photoproduction of the $J/\Psi$ at HERA
at $\sqrt{s}=200$GeV, i.e., $\xi = 8.3$
is $\sigma_{tot}(\gamma p \rightarrow
J/\Psi p) \sim 90 $nb, i.e., we predict the five-fold increase
from FNAL to HERA.
Theoretical calculations of the forward
production cross section are more straightforward and are free
of uncertainties with the energy dependence of the diffraction
slope, and it is very much desirable that the experimental data
are presented in the form of $d\sigma/dt|_{t=0}$. More detailed
comparison with the experiment will be presented elsewhere.

    In Fig.~5 we present our predictions for the effective
intercept
$$
\Delta_{eff}(\xi,Q^{2})={1\over 2}\cdot
{\partial\log (d\sigma/dt)|_{t=0})
\over \partial \xi}
$$
for the transverse and longitudinal polarizations of the $J/\Psi$
and $\rho^{0}$. The rise of $\Delta_{eff}(\xi,Q^{2})$ with
energy and $Q^{2}$ is an interplay
of the BFKL perturbative and the nonperturbative cross section.
On the other hand, for the $J/\Psi$ and the longitudinally
polarized $\rho^{0}$, at $Q^{2}\approx C^{2}/r_{\Delta}^{2}
\sim (100-200)$ GeV$^{2}$  we
predict the precocious BFKL behaviour $\Delta_{eff}(\xi,Q^{2})
\approx \Delta_{\Pom}$. Such a behaviour of
$\Delta_{eff}(\xi,Q^{2})$ is a very definitive prediction
of our approach. Because of a large value of $C_{T}(\rho^{0})$,
in the transverse $\rho^{0}$ production the magic scanning
radius $r_{S}\sim r_{\Delta}$ and $\Delta_{eff}(\xi,Q^{2})
\approx \Delta_{\Pom}$ are not attainable even at $Q^{2}\sim
200$GeV$^{2}$. This prediction is difficult to test, though,
because $\sigma_{T} \ll \sigma_{L}$. It is interesting that
one has $r_{S}\sim r_{\Delta}$ also for the quasireal
photoproduction
of the $\Upsilon(1S)$, which reaction can even be favoured for
the higher flux of quasireal photons. For the $\rho^{0}$ and
$J/\Psi$ production at large $Q^{2}$, the kinematical range
of HERA corresponds to $\exp(-\xi)\approx x \gsim 10^{-5}\cdot
(Q^{2}/{\rm GeV}^{2})$, in which region the unitarization effects
can still be neglected. We emphasize that precisely the same
approach with the same dipole cross section, gives a very good
description of the HERA data on $F_{2}(x,Q^{2})$ [15].

Within our lightcone formalism, the dipole cross section is a
universal quantity, and all
the dependence on the process is contained in wave functions of
the initial and secondary particle. The point we wish
to make is that once the different production processes
$\gamma^{*}p\rightarrow V_{i}\,p$ are studied at values $Q_{i}^{2}$
so arranged as to have the same scanning radius $r_{S}$
Eq.~(\ref{eq:4}), then the production cross sections
will exhibit identical energy dependence. One must compare
the cross sections at energy  $\nu_{i}$  corresponding to
the same rapidity $\xi$. For instance, we predict identical
energy ($\xi$) dependence of the real photoproduction of the
$J/\Psi$ and of the virtual photoproduction of the longitudinal
and transverse $\rho^{0}$ at $Q^{2}\approx 20$GeV$^{2}$ and
$\approx 30$GeV$^{2}$, respectively.
This rescaling from one vector meson to another constitutes an
important cross-check of the whole formalism.

More predictions can be made, which are specific of the small
gluon correlation radius $R_{c}=0.3$f. Namely for this reason,
the perturbative BFKL cross section $\sigma^{(pt)}(\xi,r)$ is
concentrated at smaller $r$ than the $\sigma^{(npt)}(r)$, and
the onset of the dominance of the BFKL
cross section is followed by a change of the $Q^{2}$ dependence
of the forward production cross section. In Fig.~6 we present
our predictions for the energy dependence of
$$
D_{T}=(m_{V}^{2}+Q^{2})^{4}\cdot\left. {d\sigma_{T}\over dt}
\right|_{t=0},~~
D_{L}=(m_{V}^{2}+Q^{2})^{4}\cdot{m_{V}^{2}\over Q^{2}}\cdot
\left. {d\sigma_{L}\over dt}
\right|_{t=0}, ~~
R={m_{V}^{2} \over Q^{2}}{d\sigma(\gamma^{*}_{L}\rightarrow V_{L})
{L}\over d\sigma(\gamma^{*}_{T}\rightarrow V_{T})}\,.
$$
The gradual rise of $D_{L,T}$ with $Q^{2}$ comes predominantly
from the rising density of gluons, see Eq.~(\ref{eq:9}). Notice,
that the ratio $d\sigma_{L}/d\sigma_{T}$ rises much
slower than $Q^{2}/m_{V}^{2}$ and exhibits a
substantial energy dependence
at fixed $Q^{2}$, which can be tested at HERA.
We strongly advocate studying the $Q^{2}$ dependence at fixed
rapidity $\xi$ rather than the fixed energy $\nu$.

Eqs.~(\ref{eq:9}-\ref{eq:11}) show that the vector-meson production
amplitude measures the dipole cross section
$\sigma(\xi,r_{S})$ and
the gluon structure function of the proton. We wish to emphasize
that this measurement is exceedingly sensitive to the
$r_{S}-Q^{2}$ relationship. Typically, one will probe the gluon
structure function $G(\xi,Q^{2})$ at the factorization scale
$q^{2} \approx \tau (Q^{2}+m_{V}^{2})$, where
\beq
\tau_{T,L}\approx {A\over C_{T,L}^{2}}
\label{eq:16}
\endeq
In the range of $Q^{2} \sim (10-100)$GeV$^{2}$ of the practical
interest, we find $\tau_{T,L}(J/\Psi) \sim 0.2$,
$\tau_{L}(\rho^{0}) \sim 0.15$ and
$\tau_{T}(\rho^{0}) \sim (0.07-0.1)$.
The emergence of such a dramatically small rescaling
coefficients $\tau_{T,L}$ is a consequence of colour
transparency. It was overlooked in [27].

   A brief commment on the production of the $2S$ radial excitations
$\rho'$ and $\Psi'$ is in order. The wave function of the $2S$
state has a node, because of which the corresponding weight
function $W_{T,L}(Q^{2},r^{2})$  changes the sign at $r\sim R_{V}$.
The resulting cancellations
of the $r>R_{V}$ and $r<R_{V}$  contributions to the production
amplitude (the node effect) lead to a strong suppression of
the $\rho'/\rho^{0}$ and $\Psi'/(J/\Psi)$ production ratio [7-11].
With increasing energy and
with the onset of the dominance of $\sigma^{(pt)}(\xi,r)$, which
is concentrated at smaller $r$, the node effect will decrease. For
instance, in the real photoproduction
the $\Psi'/(J/\Psi)$ ratio is expected to increase by
the factor $\sim 3$ from the CERN/FNAL energies to the highest
energies available at HERA [28]. The interesting possibility [28]
is that for the radially excited light mesons ($\phi',\,\omega',\,
\rho'$), in the real photoproduction
the $V'/V$ ratio may exhibit the anomalous,
nonmonotonic energy dependence, when this ratio first decreases
with energy, and then starts increasing.
In the opposite to
this, at large $Q^{2}$ such that $r_{S} \ll R_{V}$, the
node effect disappears, and we expect $\rho'/\rho^{0}\, \sim  \,
\Psi'/(J/\Psi)\, \sim 1$ over the whole energy range.
\medskip\\
{\Large \bf Conclusions:\\}
The purpose of this paper was to examine how the dipole cross
section $\sigma(\xi,r)$ is scanned in the virtual photoproduction
of vector mesons. The BFKL dipole cross section, complemented
by the nonperturbative dipole cross section at large $r$, gives
the unified description of the photoproduction processes in the
whole range of $\nu$ and $Q^{2}$, including the real photoproduction.
The same dipole cross section gives a good description [14] of the
proton structure function measured at HERA.
Changing $Q^{2}$, one  can probe the energy dependence of
$\sigma(\xi,r)$ in a broad range of radii from the
nonperturbative region of $r\sim 1$f down to the perturbative
domain of $r\ll 1$f. One can zoom at the magic radius $r_{\Delta}$
and determine at HERA the pomeron intercept $\Delta_{\Pom}$,
measuring
the cross section of elastic $\rho^{0},J/\Psi$ production at
at $Q^{2}\sim(100-200)$GeV$^{2}$ and the quasireal photoproduction
of the $\Upsilon(1S)$. We argued that this determination of
$\Delta_{\Pom}$ is not affected by the unitarity corrections.
We make a strong point that it is the dipole size $r$
rather than the photon's virtuality $Q^{2}$, which controls
the energy dependence of diffractive amplitudes: the value
of $Q^{2}$ needed to scan the dipole cross section at the magic
radius $r_{\Delta}$ strongly varies from the process to process,
c.f. the discussion in [15]. We predict very specific variation
with energy of the $Q^{2}$ dependence of elastic production cross
section and of the $L/T$ ratio, which derive from the small
correlation radius for perturbative gluons $R_{c}\sim 0.3$f.
Elastic production is found to probe the gluon structure
function of the proton at an anomalously small factorization
scale.
\pagebreak\\

{\bf Figure captions:}
\begin{itemize}

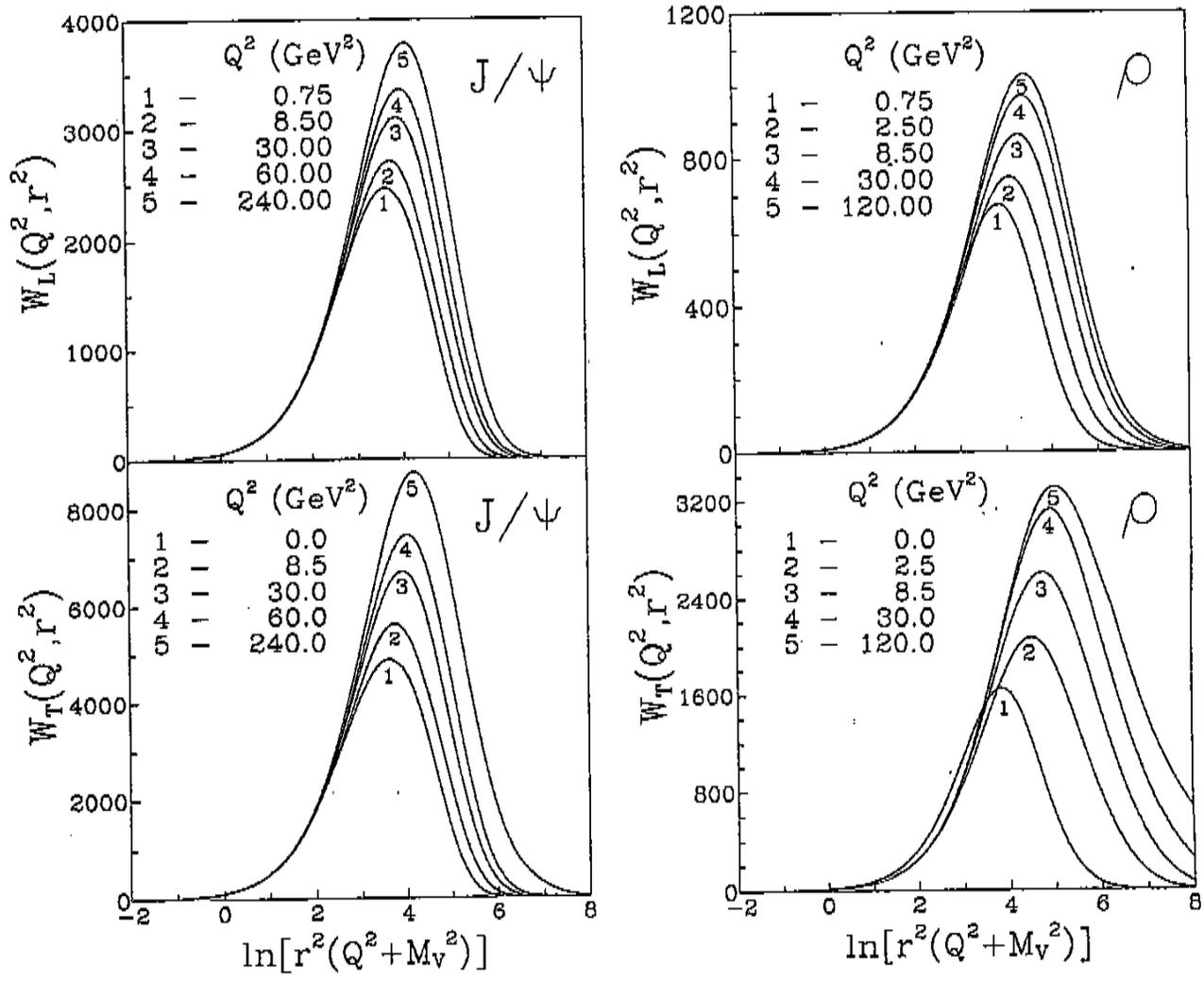
\item[Fig.~1]
{}~- Weight functions $W_{T,L}(Q^{2},r^{2})$ as a function of
a variable $y=r^{2}(Q^{2}+m_{V}^{2})$.

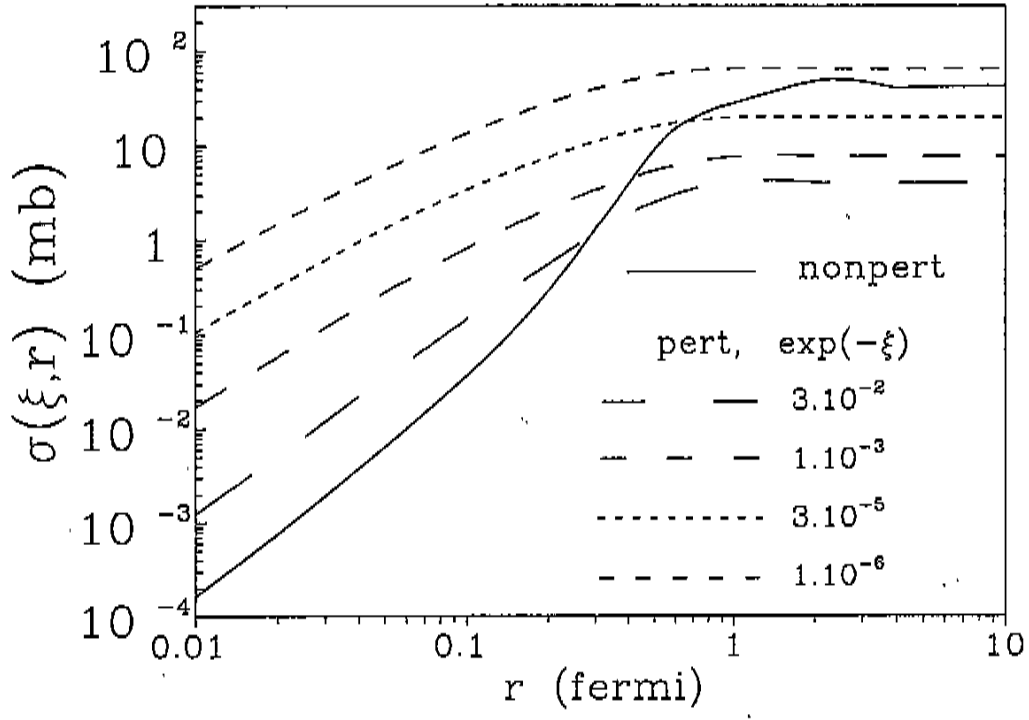
\item[Fig.~2]
{}~- Decomposition of the dipole cross section into the BFKL
perturbative $\sigma^{(pt)}(\xi,r)$ and the nonperturbative
$\sigma^{(npt)}(r)$ components. The growth of the BFKL cross
section with energy (rapidity) is shown.

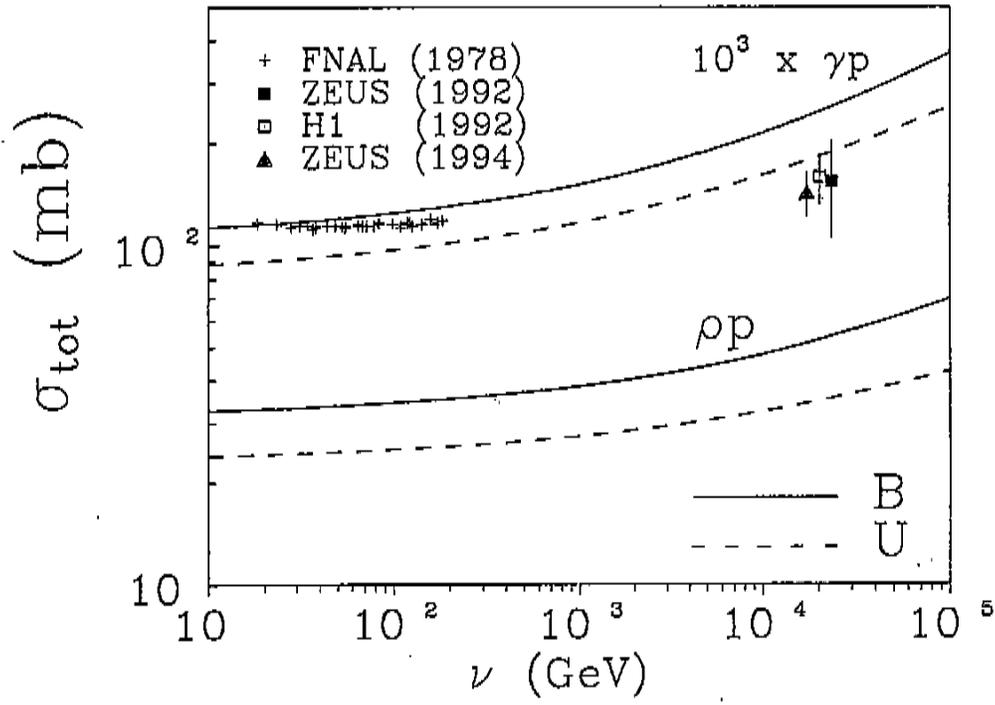
\item[Fig.~3]
{}~ - The predicted energy dependence of the (B) bare
and (U) unitarized total $\rho^{0}p$ and
real photoabsorption cross
section. The data shown are from the HERA [6] and FNAL [21]
experiments.  \\

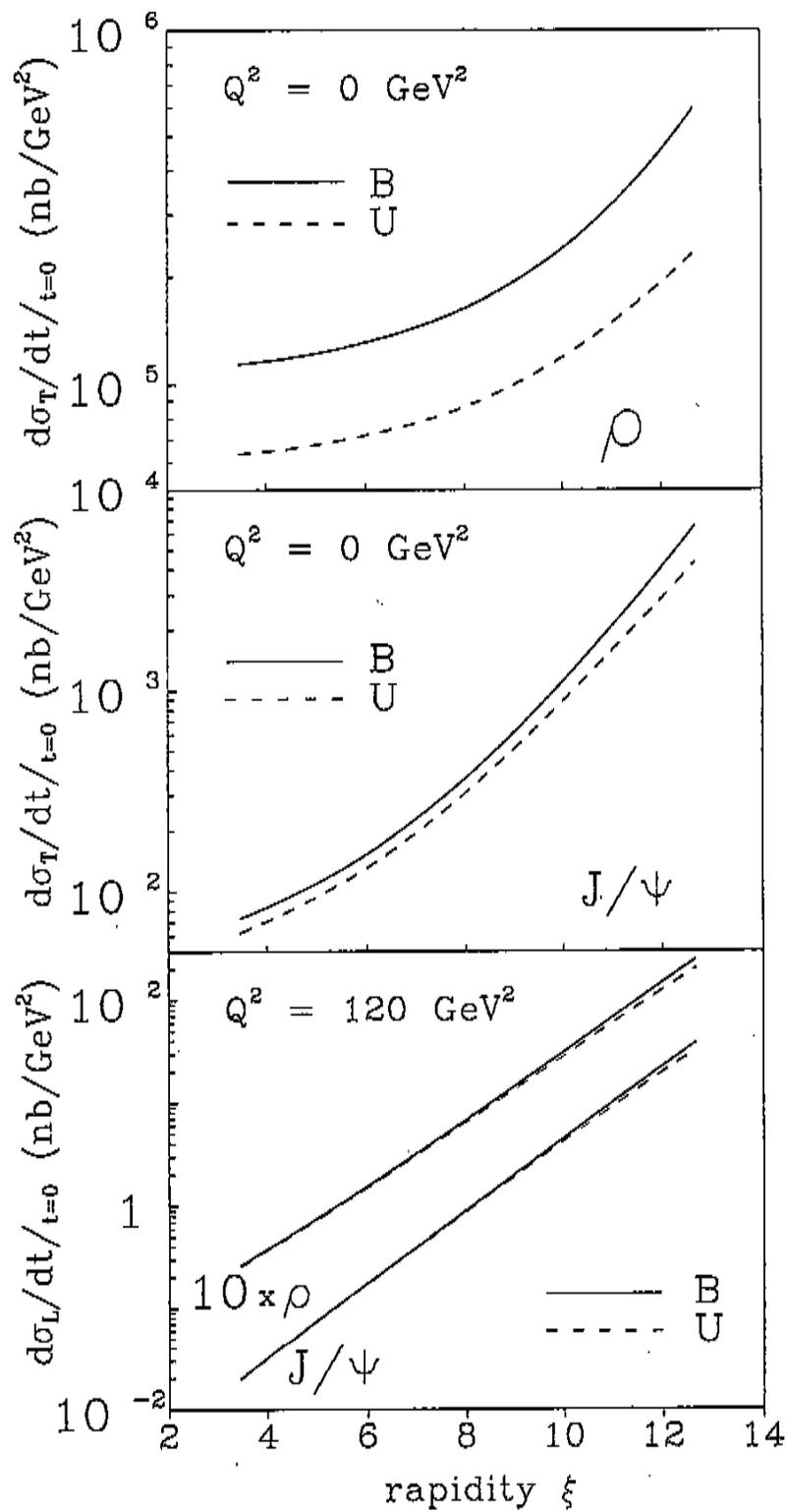
\item[Fig.~4]
{}~ - The predicted energy (rapidity) dependence of the (B) bare and
(U) unitarized differential cross section of forward
real and virtual photoproduction of the $\rho^{0}$ and $J/\Psi$.

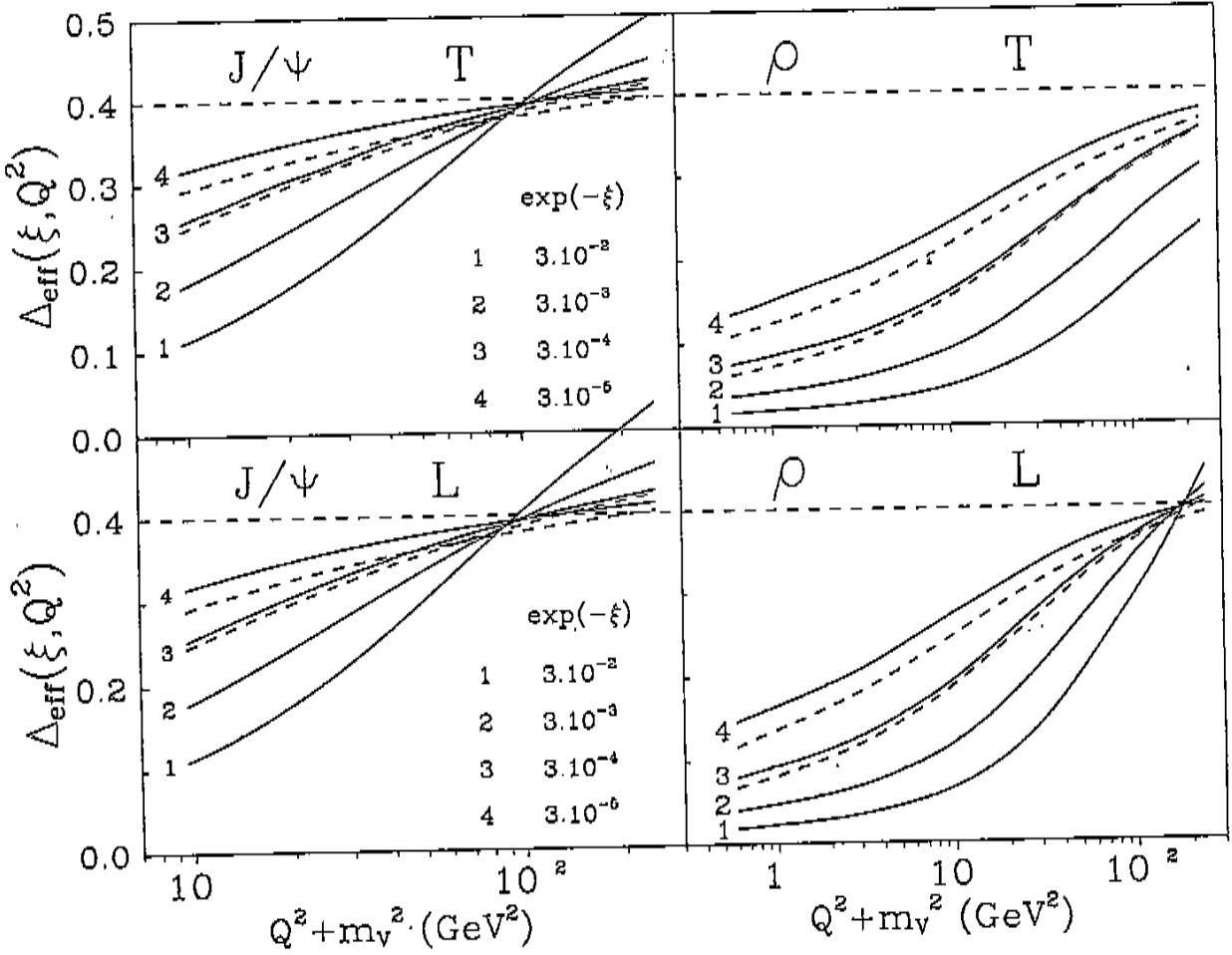
\item[Fig.~5]
{}~ - The energy (rapidity) and $Q^{2}$ dependence of the effective
intercept $\Delta_{eff}(\xi,Q^{2})$ for the forward production
of the transverse and longitudinal $\rho^{0}$ and $J/\Psi$.

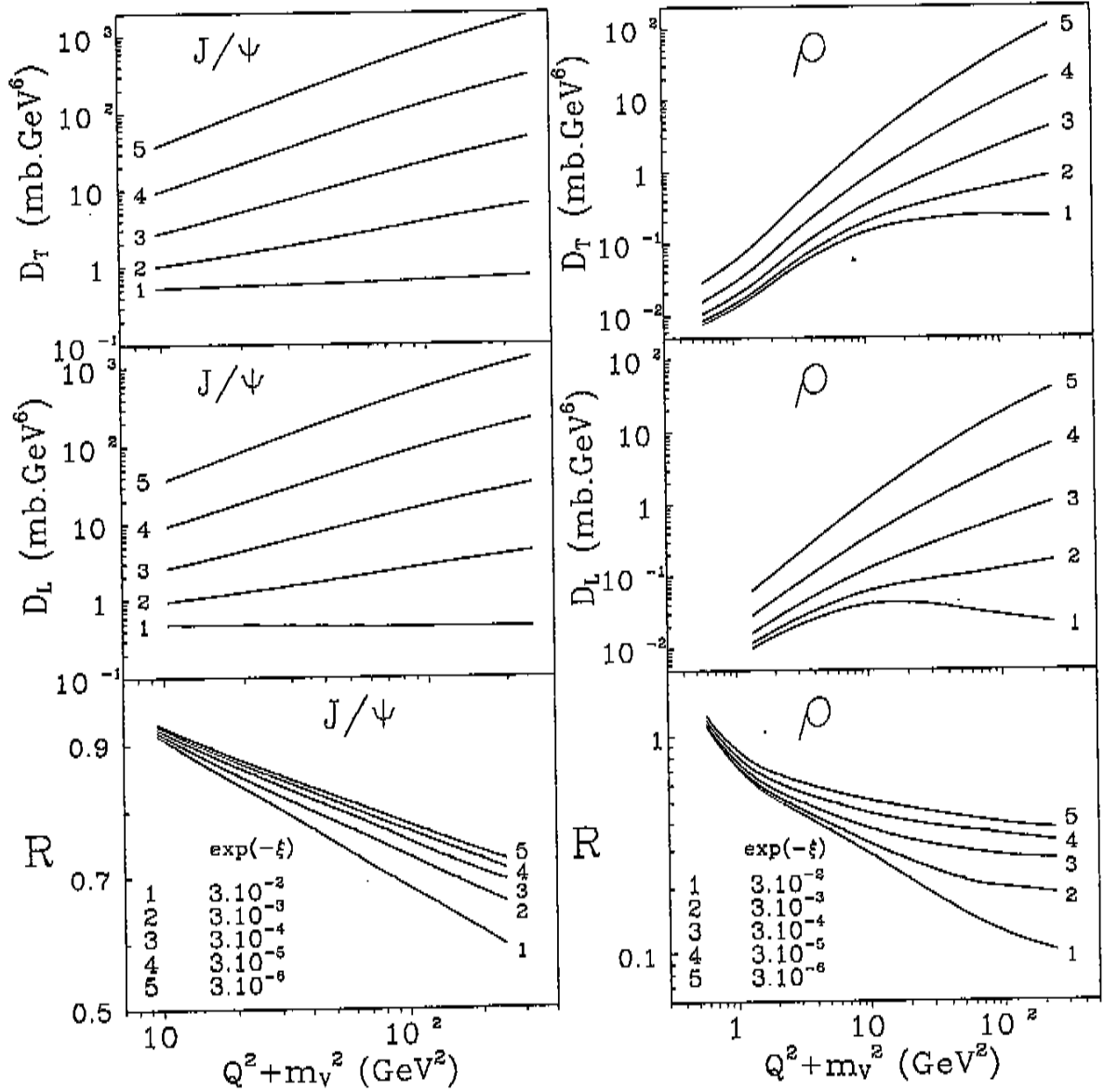
\item[Fig.~6]
{}~ - Predictions for the $Q^{2}$ dependence
of the differential cross
section of forward production and the ratio of the
longitudinal and transverse cross sections. Shown are the
quantities $D_{T,L}$ and $R$ defined in the text.
\end{itemize}
\end{document}